\documentclass[runningheads]{llncs}
\usepackage[T1]{fontenc}
\usepackage{graphicx}
\usepackage{booktabs}
\usepackage{amsmath}
\usepackage{float}
\usepackage{hyperref}
\usepackage{listings}
\usepackage{xcolor}
\usepackage{color}

\begin{document}
\title{Statistical Model Checking of the Keynes+Schumpeter Model: A Transient Sensitivity Analysis of a Macroeconomic ABM}

\titlerunning{Statistical Model Checking of the K+S Model}
\author{
  Stefano Blando\inst{1}\orcidID{0009-0007-0523-6855} \and
  Giorgio Fagiolo\inst{1}\orcidID{0000-0001-5355-3352} \and
  Mauro Napoletano\inst{2}\orcidID{0000-0002-8128-6674} \and
  Tania Treibich\inst{3}\orcidID{0000-0002-9679-1905} \and
  Andrea Vandin*\inst{1}\orcidID{0000-0002-2606-7241}
}

\authorrunning{S. Blando et al.}

\institute{
  Institute of Economics and L'EMbeDS\\
  Sant'Anna School of Advanced Studies, Pisa, Italy\\
  \email{name.surname@santannapisa.it} \and
  GREDEG, CNRS, Universit\'e C\^ote d'Azur\\
  Sophia Antipolis, France\\
  \email{mauro.napoletano@univ-cotedazur.fr} \and
  Maastricht University\\
  Maastricht, The Netherlands\\
  \email{t.treibich@maastrichtuniversity.nl}
}

\maketitle
%
\begin{abstract}
Agent-based models (ABMs) are increasingly used in macroeconomics, but their analysis still often relies on ad hoc
Monte Carlo campaigns with heterogeneous statistical effort across parameter settings. We show how statistical model
checking (SMC), implemented through MultiVeStA, can provide a principled analysis layer for a realistic macroeconomic
ABM without rewriting the simulator in a dedicated formalism. Our case study is the heuristic-switching
Keynes+Schumpeter (K+S) model, analysed through a transient sensitivity campaign over one-parameter sweeps, two macro
observables (unemployment and GDP growth), and one auxiliary micro-level probe (market share) on the post-warmup
phase of a 600-step horizon. The analysis is driven by reusable temporal queries, observable-specific precision
targets, and confidence-based stopping rules that automatically determine the simulation effort required by each
configuration. Results show a clear contrast across parameter families: macro-financial and structural sweeps produce
the strongest transient effects, whereas several heuristic-rule sweeps remain much weaker under the same precision
policy. More broadly, the paper shows that SMC can support reproducible and informative quantitative analysis of
substantively rich economic ABMs, while making uncertainty estimates and simulation cost explicit parts of the
reported results.

\end{abstract}

\section{Introduction}
\label{sec:intro}
Macroeconomic agent-based models (ABMs) are increasingly used to study economies in which aggregate
dynamics emerge from decentralized interactions among heterogeneous agents~\cite{farmer2009economy,kirman2011complex}.
They are especially useful when innovation, financial constraints, and other micro-level adaptation processes shape
macroeconomic outcomes. Within agent-based computational economics, this perspective has become a standard way of
representing macroeconomic dynamics beyond representative-agent and equilibrium-based
abstractions~\cite{tesfatsion2006handbook,dawid2018agent}, and can also be read in terms of economic multi-agent
systems~\cite{wooldridge2009introduction}. Among the best-known examples, the Keynes+Schumpeter family combines
Schumpeterian innovation dynamics with Keynesian demand interactions~\cite{dosi2010schumpeter,dosi2013income} and has
been used to study business cycles, unemployment, innovation, growth, and policy interactions in a unified
computational setting~\cite{dosi2015fiscal,dosi2017micro,dosi2020rational,fagiolo2017macroeconomic,caiani2016benchmark}.

The same features that make these models substantively interesting also make them difficult to analyse. Their behavior
is stochastic, path dependent, and intrinsically dynamic: shocks and decentralized interactions propagate through the
economy, so both micro and macro outcomes are distributions over repeated runs rather than deterministic trajectories.
Evidence is therefore obtained through simulation campaigns. In practice, however, ABM output analysis still often
relies on visual inspection and ad hoc choices about replication effort, warm-up, and fixed Monte Carlo budgets. The
statistical effort effectively spent on each parameter value and observable can therefore vary across settings, making
it harder to separate genuine behavioral differences
from calibration choices and residual simulation noise~\cite{Richiardi2006Protocol,Lee2015OutputAnalysis}, while also
weakening the transparency and comparability of cross-sweep
conclusions~\cite{fagiolo2007critical,Secchi2017,guerini2017validation,fagiolo2019validation,lamperti2018agent,Vandin2022,tieleman2022validation}.

This paper addresses that methodological gap using statistical model checking (SMC). Here, SMC imposes a common
inferential discipline on simulation-based sensitivity analysis instead of relying on fixed simulation budgets whose
effective precision varies across settings. For the comparative transient analysis considered here, this is preferable
to fixed-budget Monte Carlo because the inferential target is fixed in advance while the number of runs adapts to the
observable and parameter setting under study. We implement that perspective through MultiVeStA, a statistical analyzer
for black-box discrete-event simulators that estimates temporal properties under user-defined confidence and precision
targets while automatically determining how many runs are needed to satisfy them~\cite{sebastio2013multivesta,GilmoreRV17,Vandin2022}. Operationally, the analyst specifies the target precision in advance, and MultiVeStA keeps generating independent simulation runs until that target is met. The aim is not to propose a generic replacement for standard simulation practice, but to show that this disciplined style of analysis can be instantiated on a demanding macroeconomic model without rewriting the simulator in a special-purpose formalism. A related perspective has recently been explored in a companion Island-Model case study~\cite{blando2026island}.

Our concrete application is the heuristic-switching Keynes+Schumpeter variant studied by Dosi et al.~\cite{dosi2020rational}, rather than the baseline K+S configurations of Dosi et al.~\cite{dosi2010schumpeter,dosi2013income}. We use this variant as a benchmark because it combines heterogeneous firms, innovation and imitation dynamics, endogenous credit conditions, and expectation formation through competing forecasting heuristics.

The empirical design is a transient counterfactual sensitivity campaign built from multiple one-parameter sweeps. We focus on three
observables, \texttt{UNEMPL}, \texttt{GDP\_GROWTH}, and \texttt{MARKET\_SHARE1}, over the post-warmup portion of a
600-step simulation horizon. \texttt{UNEMPL} and \texttt{GDP\_GROWTH} capture the main macro outcomes, while
\texttt{MARKET\_SHARE1} serves as a tractable micro-level probe rather than as a full sectoral concentration index.
The exercise documents a reproducible statistical workflow for a non-trivial macroeconomic ABM and reports the resulting transient sensitivity profiles together with the computational effort required under common precision targets.

The contribution is not a new SMC algorithm but a case study showing that a common-precision workflow remains feasible and informative on a macroeconomic ABM far richer than standard methodological benchmarks.

The remainder of the paper is organized as follows. Section~\ref{sec:ks} introduces the K+S model,
Section~\ref{sec:multivesta} presents MultiVeStA, Section~\ref{sec:integration} describes the simulator interface,
Section~\ref{sec:setup} details the experimental design, and the final sections discuss results, limitations, and
future work.

\section{The Keynes+Schumpeter Model}
\label{sec:ks}

The Keynes+Schumpeter family combines Schumpeterian innovation dynamics and evolutionary industry dynamics with
Keynesian demand interactions in an agent-based macroeconomic setting, and is now a well-established reference point
in the ACE literature on business cycles, policy, and endogenous growth~\cite{dosi2010schumpeter,dosi2013income,dosi2015fiscal,dosi2017micro}. It is also
widely discussed as a benchmark family in the broader macro-ABM literature~\cite{dosi2020rational,fagiolo2017macroeconomic,caiani2016benchmark} and can
be read within a longer tradition of evolutionary and history-friendly model building~\cite{capone2019history}. The
family draws on evolutionary theories of innovation and Schumpeterian growth, including learning, search, and
creative-destruction processes~\cite{nelson1982evolutionary,arrow1962economic,aghion1992model}.
The specific variant used in this paper is the model of heuristic expectation switching studied by Dosi et
al.~\cite{dosi2020rational}. It features two production sectors, heterogeneous firms, endogenous entry and exit,
endogenous loan allocation across banks and firms, and a labor market with aggregate employment adjustment. At the same time, the model retains
a strong behavioral dimension: consumption-good firms form demand expectations using alternative heuristics and may
switch among them over time.

The model is organized around a capital-good sector, a consumption-good sector, households, banks, and
the public sector~\cite{dosi2010schumpeter,dosi2013income,dosi2015fiscal}. Capital-good firms innovate and imitate to
supply heterogeneous machines; consumption-good firms decide output, investment, and prices under heterogeneous
technological and financial conditions; aggregate demand, employment, and GDP growth emerge from decentralized firm
decisions. Unemployment, GDP growth, and firm-level market share are therefore jointly shaped by innovation, credit
conditions, and fiscal policy, so that even one-dimensional parameter changes can produce non-trivial transient
effects~\cite{dosi2010schumpeter,dosi2013income,dosi2015fiscal,dosi2017micro}. This is why the campaign does not
restrict attention to heuristic-switching parameters alone: several sweeps target policy, financial, and structural
parameters that shape the environment in which expectation formation takes place.

The expectation module is central for our study because it provides both the main behavioral novelty of the selected
variant and a natural interface between economic interpretation and statistical sensitivity analysis. In the switching
version of the model, consumption-good firms can rely on a naive rule, the simplest parameter-free benchmark inherited from the non-switching variants, adaptive expectations, weak trend following,
strong trend following, or an anchor-and-adjustment rule~\cite{dosi2020rational,tversky1974judgment}. The relative attractiveness of these
heuristics depends on recent forecasting performance, and their population shares evolve through a switching mechanism
governed by interpretable parameters such as the intensity of choice, persistence in rule adoption, and memory of past
performance~\cite{dosi2020rational,anufriev2012evolutionary,brock1997rational}. Several coefficients internal to the individual heuristics can also be varied directly, linking changes in
belief formation to changes in macroeconomic and competitive dynamics. In this sense, the switching block can also be
read against the broader traditions of bounded rationality and heterogeneous expectations in economics, and against laboratory evidence on expectation-feedback dynamics~\cite{simon1955behavioral,hommes2006heterogeneous,hommes2011heterogeneous,heemeijer2009price,anufriev2013evolutionary}.

\section{MultiVeStA as an Analysis Layer}
\label{sec:multivesta}

This section briefly introduces the estimation-oriented SMC perspective used in the paper and
explains why MultiVeStA is the relevant analysis engine for our K+S case study.

Statistical model checking estimates properties of stochastic systems from repeated simulation
runs, with statistical guarantees whose tightness depends on user-chosen confidence and precision
parameters~\cite{younes2002probabilistic,legay2010statistical,agha2018survey,DBLP:books/daglib/0007403-2,DBLP:books/daglib/0020348}.
It covers both hypothesis-testing and estimation-oriented analyses~\cite{agha2018survey,wald1945sequential}; the
present paper is firmly in the second category, estimating transient expectation profiles under controlled precision
targets rather than certifying a Boolean property. This is especially appealing for ABMs, whose state spaces are
typically too large and whose implementations are often available only as executable simulators.

MultiVeStA is a statistical model checking framework for stochastic simulators that can be controlled as black
boxes~\cite{sebastio2013multivesta,Vandin2022,GilmoreRV17}. Rather than requiring an explicit state-space
model, it relies on repeated simulation runs and on a lightweight interaction protocol through which the analysis
engine can reset the simulator, advance it step by step, and read user-defined observables. This black-box style of
integration has been progressively extended across different simulation settings, including economic ABMs~\cite{Vandin2022}, Python-based ABMs~\cite{DBLP:conf/vecos/Vandin24}, and NetLogo models~\cite{DBLP:journals/corr/abs-2509-10977}, with a corresponding emphasis on distributed statistical execution and reusable analysis pipelines~\cite{DBLP:conf/hpcs/PianiniSV14,Vandin2022}.

The properties analysed in MultiVeStA are specified in MultiQuaTEx, a query language geared toward expectations and
other statistical summaries over simulation traces~\cite{sebastio2013multivesta,GilmoreRV17}. In the transient setting considered here, a query asks for the expected value of an observable at a family of time points, so
MultiVeStA estimates a trajectory of expectations rather than a single end-of-run scalar. It keeps generating
independent runs until the requested confidence-interval half-width falls below a user-provided threshold, so the
stopping rule is attached to the estimate rather than to a fixed simulation budget~\cite{Vandin2022}. This separates
three ingredients that are often entangled in ad hoc simulation campaigns: the simulator generates traces, the query
defines the quantity to be estimated, and the statistical engine decides how much simulation effort is required.
The query, the precision target, and the stopping rule are therefore part of the inferential design of the experiment.

This mode of operation is particularly useful for sensitivity analysis. Once a precision target is fixed, different
parameter values and different observables can be compared under a common statistical discipline, and the computational
cost required to reach that precision can be reported explicitly. The combination of automated stopping, black-box
compatibility, and parallel execution is what makes MultiVeStA a suitable analysis engine for the present K+S case
study.

\section{Integrating MultiVeStA with the K+S Model}
\label{sec:integration}
\subsection{Interface and Exposed Observables}
\label{sec:integration-interface}

To make the K+S simulator analyzable by MultiVeStA, we instrument the existing C++ implementation with a thin bridge
layer around the existing code in \texttt{main.cpp}. Rather than refactoring the model into a separate object-oriented
wrapper, the integration is implemented through a dedicated entry point, \texttt{main\_mv}, which reads commands from
standard input. At the level of interaction, this yields the three operations needed by the statistical engine:
resetting the simulator with a fresh seed, advancing the simulation by one step, and evaluating named observables on
the current state. This design keeps the economic model and the statistical analysis conceptually separate: the
simulator remains responsible for the underlying dynamics, while MultiVeStA orchestrates the repeated iid executions
needed for estimation.

The interface is deliberately minimal. MultiVeStA does not need direct access to the internal data structures of the
model, nor does it require the simulator to be rewritten as an explicit stochastic state-space system. It only needs a
repeatable command protocol through which a run can be initialized, advanced, and queried. This is what makes the
integration attractive for a legacy ABM code base such as K+S.

\begin{lstlisting}[language=C++,caption={Excerpt of the command loop used to connect the simulator to MultiVeStA.},label={lst:mv-interface}]
while (shall_continue) {
  std::string read;
  std::cin >> read;

  if (read == "next") {
    t++;
    performOneStep();
  } else if (read.rfind("reset", 0) == 0) {
    reset_once_at_beginning();
    reset_once_per_parametrization();
    reset_once_per_simulation(mv_seed_long);
    t = 1;
    performOneStep();
  } else {
    std::cout << "OUTPUTMV:" << eval(read) << std::endl;
  }
}
\end{lstlisting}

The integration exposes the paper's observables through the \texttt{eval} function: unemployment, GDP growth, and
firm-level market shares. Observations are returned through the \texttt{OUTPUTMV:} channel shown in
Listing~\ref{lst:mv-interface}, while the simulation clock is maintained by the stepwise execution loop. MultiQuaTEx
can therefore refer to the built-in step counter \texttt{steps} separately from the model-specific observables.

\begin{lstlisting}[language=C++,caption={Observable mapping exposed by the current implementation.},label={lst:evalobs}]
double eval(std::string obs) {
  if (obs == "GDP") return GDP(1);
  else if (obs == "GDP_GROWTH") return GDP_g2;
  else if (obs == "UNEMPL") return U(1);
  else if (obs.find("MARKET_SHARE", 0) != string::npos) {
    int firm = std::stoi(obs.substr(12));
    return f2(1, firm);
  }
  return -1;
}
\end{lstlisting}

\subsection{From Queries to Statistical Estimates}
\label{sec:integration-workflow}

Once the simulator is exposed through this interface, MultiVeStA can execute the full analysis loop automatically. For
each parameter setting, it launches the binary with the flags \texttt{-experimentMV} and \texttt{-numMCexpMV}, which
select one sweep and one specific parameter value. It then initializes the model with an independent seed, advances the
simulation step by step, evaluates the MultiQuaTEx query on the resulting traces, and accumulates observations until
the requested confidence interval width has been reached. The result of a single job is therefore a mean trajectory
equipped with confidence intervals and with the number of simulations required to obtain it.

Initialization is split into a part executed once at the beginning, a part executed once per parametrization, and a
part executed once per simulation run. This mirrors the structure of the sensitivity campaign: some settings define
the sweep configuration, while others must be freshly drawn for each iid execution.

\begin{lstlisting}[caption={Transient MultiQuaTEx pattern used in the campaign. The concrete query files only differ in the observable name.},label={lst:query}]
obsAtStep(x, obs) =
  if (s.rval("steps") == x) then s.rval(obs)
  else # obsAtStep(x, obs) fi;

eval parametric(E[obsAtStep(x, obs)], x, 101, 10, 600);
\end{lstlisting}

Operationally, a query of this form asks for the expected post-warmup trajectory of one observable
under a fixed parameter setting, so the output can be read directly as a statistically disciplined
sensitivity profile rather than as a single end-of-run average.

This separation between simulator, query, and stopping rule is central for the paper: the economic model defines the
dynamics, the interface defines what can be observed, the query defines which temporal property is estimated, and
MultiVeStA determines how much simulation effort is needed. The bridge is therefore not just a technical convenience
but the point at which the economic simulator becomes amenable to a reproducible statistical
workflow~\cite{DBLP:conf/ifm/GilmoreTV14}. The next section instantiates this workflow for the concrete K+S campaign.

\section{Experimental Design}
\label{sec:setup}

This section is organized around four blocks: model configuration, query design, parameter and
observable selection, and precision calibration. Together, these choices define the statistical
protocol used for the full sensitivity campaign.

\subsection{Model Configuration}
\label{sec:config}

We analyse a C++ implementation of the K+S model compiled with
\texttt{flagEXP\_switch\,=\,1} and \texttt{flag\_OLS\,=\,0}, so the reported experiments use the
heuristic-switching configuration with OLS learning disabled. Each run lasts 600 discrete time
steps~\cite{dosi2020rational}; analysis focuses on the post-warmup window from step~101. The
campaign decomposes into independent MultiVeStA runs, one per (experiment, parameter value,
observable) combination, invoked via flags \texttt{experimentMV} and \texttt{numMCexpMV}.
Computational cost is not fixed a priori: each run stops when MultiVeStA's confidence-driven
criterion is satisfied.
For scale, the realized mean sample counts in the campaign range from about 1\,450 to about
14\,000 runs depending on the experiment-observable pair, far above the conventional fixed Monte
Carlo budgets of 50 or 100 runs often used in exploratory practice.

All analyses use $\alpha = 0.05$, 40 parallel workers, block size 30, and a common seed-of-seeds
value of 1, with no fixed simulation cap.

\subsection{Query Design}
\label{sec:query}

The transient queries used in the paper all share the same structure. Listing~\ref{lst:query} shows the generic
MultiQuaTEx pattern: for a given observable name `obs', a recursive operator waits until the simulator reaches step
$x$ and then returns the current value of that observable. The concrete query files differ only in whether
\texttt{obs} is instantiated as \texttt{UNEMPL}, \texttt{GDP\_GROWTH}, or \texttt{MARKET\_SHARE1}. The statistical
settings used to execute these queries --- in particular $\alpha = 0.05$, 40 workers, and block size 30 --- were
defined in the model-configuration block above and remain fixed throughout the campaign.

For each run, the simulator is therefore observed at steps $101, 111, 121, \ldots, 591$, yielding 50 post-warmup
evaluation points on a common temporal grid.
Throughout the paper, these steps are treated as the simulator's native discrete time units; the analysis does not
rely on an external calendar interpretation of the horizon.

Two design choices matter. First, the paper targets the full post-warmup transient rather than a small terminal
fragment of the simulation. Second, all observables are queried on the same temporal grid, which keeps the
interpretation of the figures and the comparison across experiments uniform.

The first two observables are macro variables, whereas \texttt{MARKET\_SHARE1} is a deliberately micro-level probe
used to check whether a parameter change alters competitive positions at the firm level. This common query template is
important for reproducibility because it keeps the statistical procedure fixed while changing only the quantity being
observed.

\subsection{Parameters and Observables}
\label{sec:params}

The campaign spans twelve one-dimensional sensitivity experiments, denoted by `E1' through `E12' in the paper and by
\texttt{experiment=2,3,4,7,10,\ldots,17} in the code. `E1'--`E3' target the heuristic-switching mechanism directly;
`E5'--`E8' vary coefficients internal to individual forecasting rules; and `E4', `E9'--`E12' modify the broader
policy, credit, and structural environment in which those rules operate. The sweep grids are reproduced in
Tables~\ref{tab:params-a} and~\ref{tab:params-b}. This grouping separates three economic layers of the model:
meta-switching, rule-specific behavior, and the surrounding macro-financial environment.

The three observables were selected to balance macroeconomic relevance and technical tractability. `UNEMPL' captures
aggregate labor-market conditions, `GDP\_GROWTH' captures output dynamics in rate form, and `MARKET\_SHARE1' monitors
the market share of one tracked consumption-good firm. If the goal were to measure sectoral concentration, an
aggregate index such as Herfindahl would be more appropriate. Here, instead, `MARKET\_SHARE1' answers the narrower
question of whether a parameter change shifts competitive positions at the micro level at all, without adding a new
layer of sector-wide aggregation logic to the simulator interface. Tracking the same firm across all runs is useful
not because that firm is statistically representative, but because it provides a stable probe of whether the
competitive margin reacts at the micro level when the sweep parameter changes.

\begin{table}[h!]
\caption{Heuristic-switching and forecasting-rule sweep experiments. Within-sweep baseline values in \textbf{bold}.}
\label{tab:params-a}
\small
\centering
\begin{tabular}{p{0.07\textwidth}p{0.23\textwidth}p{0.38\textwidth}p{0.24\textwidth}}
\toprule
ID & Variable & Role in the model & Sweep values \\
\midrule
E1 & $\beta$ (beta\_switch) & intensity of choice in heuristic switching & 0.0, 0.2, \textbf{0.4}, 0.6, 0.8, 1.0 \\
E2 & $\delta_s$ (delta\_switch) & inertia in heuristic adoption probabilities & 0.0, 0.25, 0.5, \textbf{0.7}, 0.9, 1.0 \\
E3 & $\eta$ (eta\_switch) & memory in heuristic performance scores & 0.0, 0.25, 0.5, \textbf{0.7}, 0.9, 1.0 \\
E5 & $\omega_{ada}$ (delta1) & coefficient of adaptive expectations (ADA/WADA rule) & 0.2, 0.4, \textbf{0.65}, 0.8, 1.0, 1.2 \\
E6 & $\omega_{wtr}$ (w3) & coefficient of weak trend-following expectations & 0.1, 0.2, \textbf{0.4}, 0.6, 0.8, 1.0 \\
E7 & $\omega_{str}$ (w4) & coefficient of strong trend-following expectations & 0.7, 1.0, \textbf{1.3}, 1.6, 2.0, 2.5 \\
E8 & $\omega_{aa}$ (laa) & weight on aggregate vs idiosyncratic growth in anchor-and-adjustment expectations & 0.0, 0.2, 0.4, \textbf{0.5}, 0.8, 1.0 \\
\bottomrule
\end{tabular}
\end{table}

\begin{table}[t]
\caption{Policy, financial, and structural sweep experiments. Within-sweep baseline values in \textbf{bold}.}
\label{tab:params-b}
\small
\centering
\begin{tabular}{p{0.07\textwidth}p{0.23\textwidth}p{0.38\textwidth}p{0.24\textwidth}}
\toprule
ID & Variable & Role in the model & Sweep values \\
\midrule
E4 & $\tau$ (aliq) & tax rate on profits and public-budget inflow & 0.0, 0.05, \textbf{0.10}, 0.15, 0.20, 0.25 \\
E9 & $\tau_b$ (credit\_multiplier) & bank reserve or capital requirement proxy & 0.04, 0.06, \textbf{0.08}, 0.10, 0.12, 0.16 \\
E10 & $\mu_{deb}$ (bankmarkup\_init) & initial bank markup on loan rates at model initialization (not a bank-size parameter) & 0.1, 0.2, \textbf{0.3}, 0.4, 0.5, 0.7 \\
E11 & $a$ (pareto\_a) & Pareto-shape parameter for firm-size heterogeneity & 0.4, 0.6, \textbf{0.8}, 1.0, 1.2, 1.5 \\
E12 & $\iota$ (theta) & inventory target or expected-stock ratio & 0.05, \textbf{0.10}, 0.15, 0.20, 0.25, 0.30 \\
\bottomrule
\end{tabular}
\end{table}

Tables~\ref{tab:params-a} and~\ref{tab:params-b} report the value grid for the varied parameter of each sweep. All
other parameters remain at the code-defined reference settings used in the corresponding initialization branch. The
relevant baseline notion in the paper is therefore the within-sweep baseline, not a single global baseline shared by
all experiments.

\subsection{Precision Parameters ($\varepsilon = \delta$)}
\label{sec:delta}

Following the calibration rationale discussed by Vandin et al.~\cite{Vandin2022}, we set the precision target
for each observable so that the confidence interval half-width $\delta$ matches the minimum effect size we aim to
distinguish across parameter settings. In the concrete experiment scripts, this is implemented by passing
observable-specific \texttt{delta-list} values to MultiVeStA: \texttt{[0.005]} for unemployment, \texttt{[0.003]} for
GDP growth, and \texttt{[0.0005]} for market share. In this sense, $\varepsilon = \delta$ plays a methodological role:
it encodes what counts as a practically meaningful separation between sensitivity curves, while also determining the
amount of simulation effort required by MultiVeStA. This $\delta$ is therefore a precision parameter on the estimated
expectation, not the spacing between adjacent parameter values in a sweep grid.

The same $\delta$ value is used for all parameter sweeps of a given observable. This avoids adapting statistical
effort from one experiment to another in ways that make comparisons harder to interpret. Fixing the target precision
at the observable level lets us report computational cost differences as an outcome of the model's behavior rather
than of analyst discretion.

\begin{table}[t]
\caption{Observable-specific precision targets used across the full campaign. Here $\delta$ denotes the target confidence-interval half-width used by MultiVeStA, not the step size between neighboring parameter values.}
\label{tab:deltas}
\small
\centering
\begin{tabular}{p{0.27\textwidth}p{0.12\textwidth}p{0.39\textwidth}}
\toprule
Observable & $\delta$ & Typical scale \\
\midrule
UNEMPL & 0.005 & $\approx 0.047$ in the reference regime \\
GDP\_GROWTH & 0.003 & $\approx 0.03$ in the reference regime \\
MARKET\_SHARE1 & 0.0005 & $\approx 0.005$--$0.006$ in the reference regime \\
\bottomrule
\end{tabular}
\end{table}

These values correspond to roughly a 10\% relative scale for all three observables. The uniform precision policy is
thus a substantive design decision of the paper: it makes the campaign reproducible, keeps statistical uncertainty
interpretable across figures, and provides a clean basis for reporting simulation costs together with economic
sensitivity patterns.
\section{Results}
\label{sec:results}

\begin{figure}[h]
\centering
\includegraphics[width=0.9\textwidth]{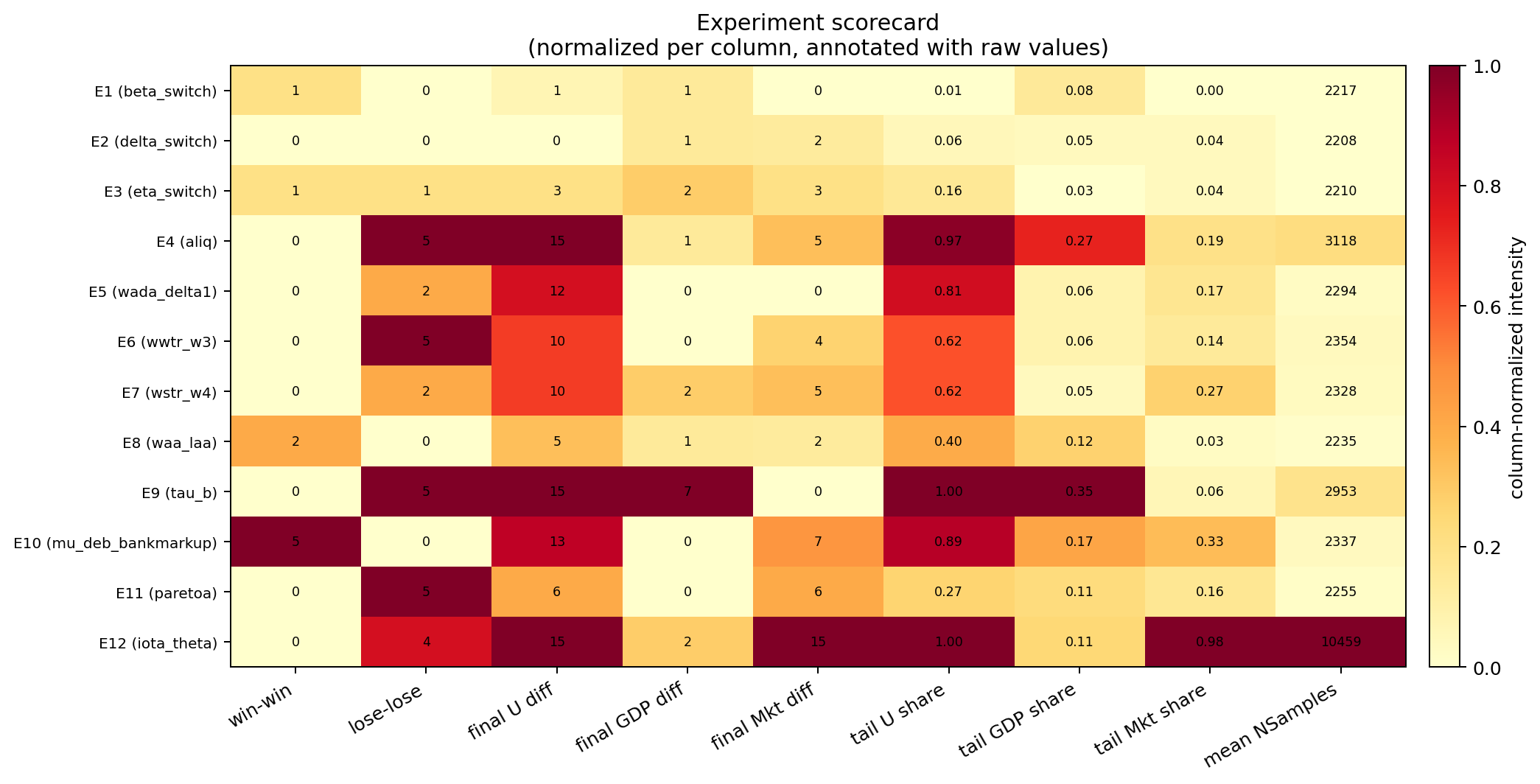}
\caption{Cross-experiment scorecard of the transient sensitivity campaign. Each row summarizes one
sweep by trade-off counts, counterfactual separation, best tail signal, and mean sample complexity
across the three observables.}
\label{fig:scorecard}
\end{figure}

We summarize the twelve one-parameter sweeps through a cross-experiment scorecard
(Fig.~\ref{fig:scorecard}) and a small set of representative plots in the body.
Appendix~\ref{sec:appendix-tables} reports the full summary tables and
Appendix~\ref{sec:appendix-figures} collects the remaining plots.

In the scorecard, trade-off counts compare each sweep point with the within-sweep baseline `p1':
`win-win' denotes higher GDP growth with lower unemployment, `lose-lose' the opposite, and
`mixed / trade-off' the remaining directional combinations. `Best tail signal' reports the most
favorable tail mean reached within a sweep, while `mean sample complexity' reports the mean number
of simulations required by MultiVeStA for that experiment-observable block.
These scorecard labels are directional summaries of the transient responses. Statistical strength is
evaluated separately through significant pairwise-separation metrics such as `Final diff.' and
`Mean tail diff. share' in the appendix tables.

The scorecard reveals a clear gradient of sensitivity. The switching parameters `E1'--`E3' produce
almost no pairwise separation under the common precision policy, `E4'--`E8' are heterogeneous, and
`E9'--`E12' contain the strongest and most persistent effects. Unemployment is the most reactive
observable: `E4', `E9', and `E12' achieve full final-step separation (15 of 15 pairs), and `E10'
reaches 13 of 15. GDP growth is harder to move, with `E9' as the strongest case,
while market share responds selectively, most notably under `E12' and `E10'. Full values appear in
the appendix tables.

\subsection{Switching Parameters (E1--E3)}
\label{sec:results-switching}

The direct switching sweeps are the weakest block in the campaign. For unemployment, $\beta$
(`E1') and $\delta_s$ (`E2') yield at most one significantly different pair at the final sampled
step; $\eta$ (`E3') is somewhat stronger but still reaches only 3 of 15 final pairs and a
tail-difference share of 0.16. GDP growth and market share are even less responsive. The
meta-level switching mechanism therefore appears locally stable over the explored ranges.
Supplementary plots for `E1'--`E3' appear in Appendix~\ref{sec:appendix-figures}.

\subsection{Fiscal Policy Parameter (E4)}
\label{sec:results-fiscal}

The tax rate $\tau$ (`E4') achieves full pairwise separation for unemployment at the final
sampled step (15 of 15 pairs) and a near-saturated tail-difference share of 0.97. The pattern is
monotone: higher tax rates raise unemployment and reduce GDP growth, yielding a ``lose-lose''
trade-off classification for all five non-baseline points relative to the zero-tax baseline.
This is consistent with fiscal policy acting as a net burden on firm profitability in this model
variant, where higher taxes reduce retained earnings and investment capacity without a
sufficiently compensating stabilization channel~\cite{dosi2015fiscal}, and makes `E4' the
clearest policy-driven unemployment case in the campaign.

\begin{figure}[h!]
\centering
\includegraphics[width=0.9\textwidth]{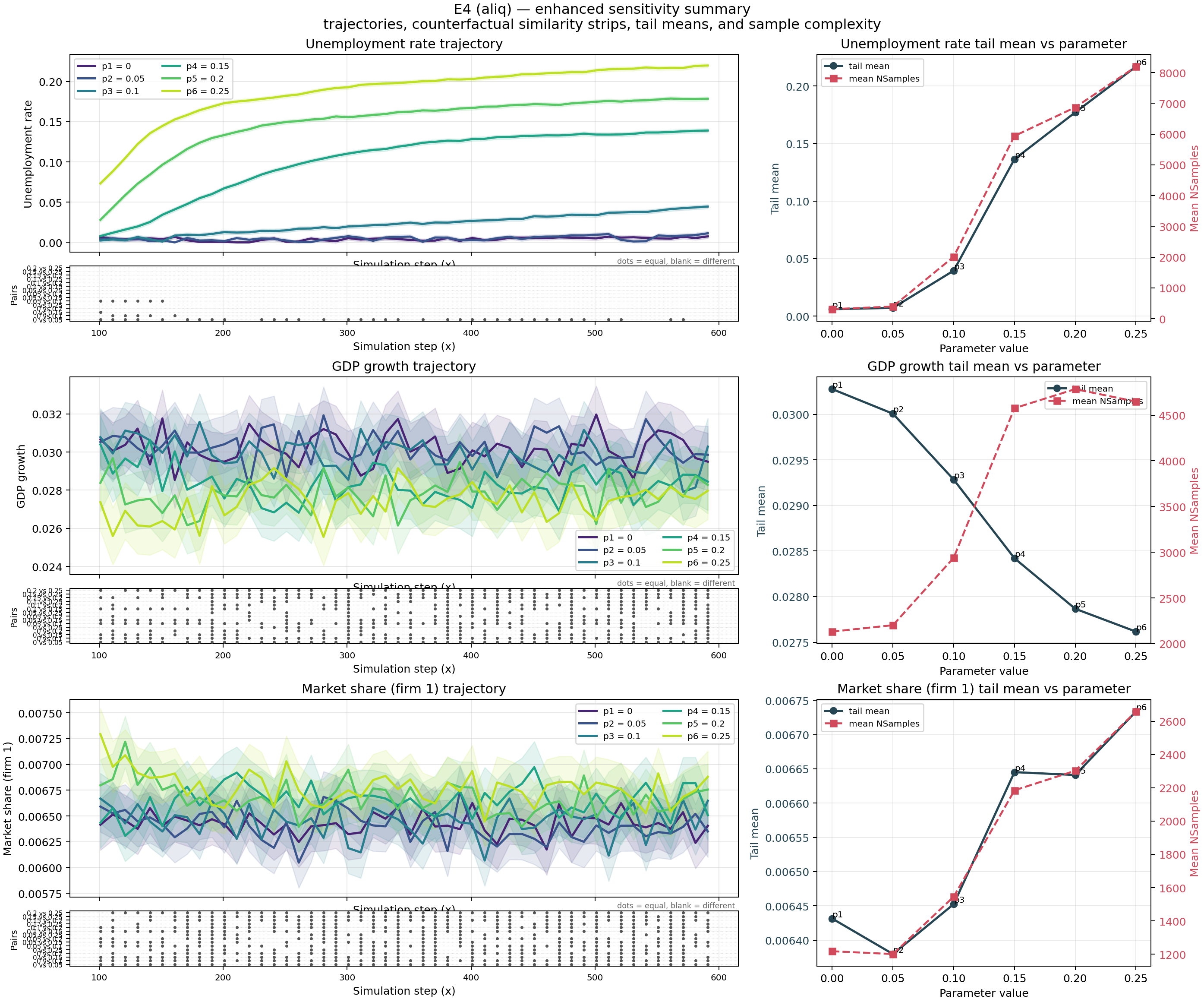}
\caption{Sensitivity summary for `E4' ($\tau$, tax rate). Unemployment bands fan out early and
monotonically, while GDP growth and market share remain more weakly separated.}
\label{fig:e4-aliq}
\end{figure}

Figure~\ref{fig:e4-aliq} shows that this effect is persistent rather than confined to a narrow
transient window: unemployment bands diverge early and remain separated, whereas GDP growth and
market share display only limited pairwise contrasts.

\subsection{Heuristic Rule Coefficients (E5--E8)}
\label{sec:results-heuristics}

The rule-internal coefficients `E5'--`E8' often separate unemployment trajectories --- `E5' reaches
12 of 15 final pairs and `E6' reaches 10 --- but remain weak or intermittent for GDP growth and
market share. Their effects are therefore selective rather than system-wide. Supplementary plots
for `E5'--`E8' are provided in Appendix~\ref{sec:appendix-figures}.

\subsection{Financial and Structural Parameters (E9--E12)}
\label{sec:results-structural}

The macro-financial and structural sweeps contain the strongest signals of the campaign: the
statistically strongest macro-level case (`E9'), the clearest directional win-win pattern (`E10'),
the statistically strongest market-share case with high computational cost (`E12'), and a weak
contrast (`E11').

\paragraph{Credit parameter $\tau_b$ (E9).}
The credit sweep is the only experiment that combines strong unemployment separation (15 of 15
final pairs, tail share 1.00) with a clearly non-trivial GDP-growth signal (7 of 15 final pairs,
tail share 0.35). All five non-baseline points are classified as ``lose-lose'' relative to the
sweep baseline, indicating that tighter credit requirements simultaneously raise unemployment and
reduce GDP growth.

\begin{figure}[h!]
\centering
\includegraphics[width=0.9\textwidth]{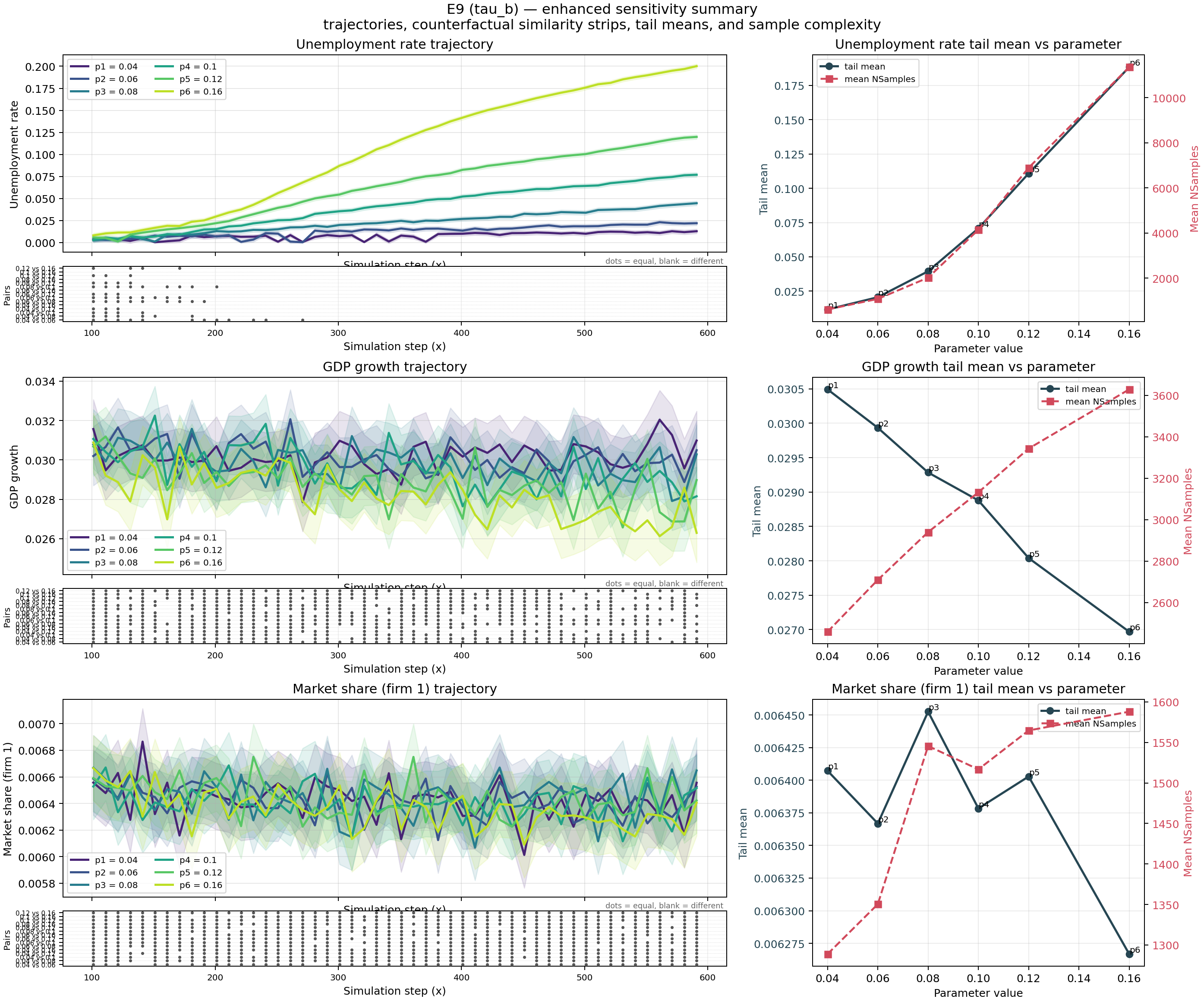}
\caption{Sensitivity summary for `E9' ($\tau_b$, credit parameter). Unemployment bands
diverge sharply, GDP growth separates more modestly, and market-share effects remain limited.}
\label{fig:e9-taub}
\end{figure}

Figure~\ref{fig:e9-taub} shows that unemployment trajectories under `E9' separate early and
persistently, while GDP-growth bands exhibit visible but narrower divergence. This is consistent
with credit constraints amplifying labor-market fluctuations more directly than aggregate output
in the short run~\cite{dosi2013income}. Market share remains largely indifferent to $\tau_b$
across the explored range.

\paragraph{Initial bank markup $\mu_{deb}$ (E10).}
The initial-bank-markup sweep is the clearest directional ``win-win'' case in the campaign. All five
non-baseline points improve the joint unemployment-GDP trade-off relative to the sweep baseline,
and 7 of 15 market-share pairs are also significantly different at the final step. The
unemployment signal is strong (13 of 15 final pairs, tail share 0.89), whereas GDP growth
improves directionally but remains only weakly separated statistically.

\begin{figure}[h]
\centering
\includegraphics[width=0.9\textwidth]{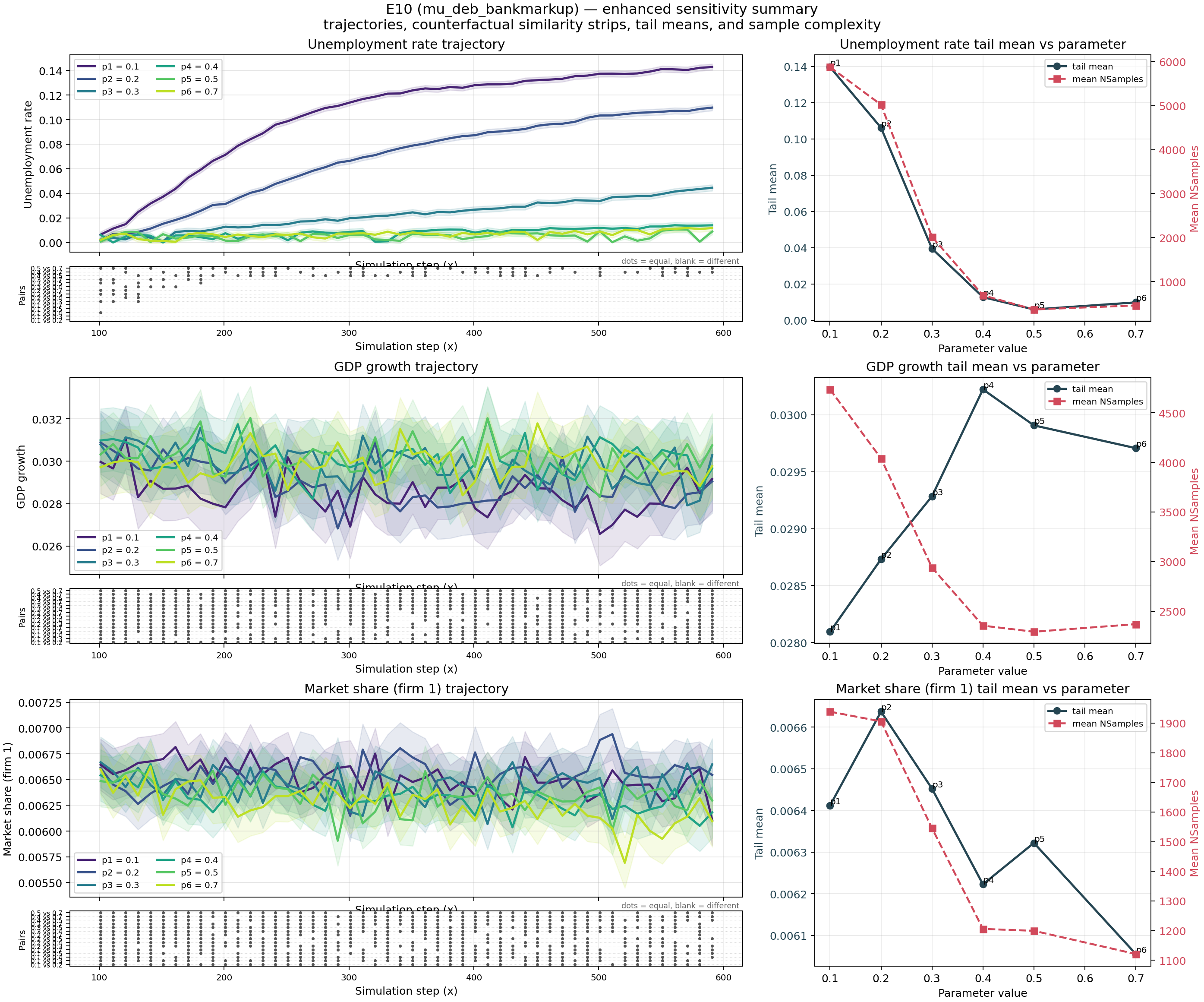}
\caption{Sensitivity summary for `E10' ($\mu_{deb}$, initial bank markup on loan rates). This is the clearest
directional win-win case: unemployment improves strongly, GDP growth shifts in a favorable
direction, and market share also moves across the sweep.}
\label{fig:e10-bankmarkup}
\end{figure}

Figure~\ref{fig:e10-bankmarkup} illustrates this directional win-win pattern: unemployment bands
separate clearly, market share also moves, and GDP growth shifts in a favorable direction without
becoming a strong statistical-separation case. The result is therefore best read as a robust
joint directional improvement rather than as evidence that GDP growth is generally easy to
separate under this campaign design.

\paragraph{Inventory expectations $\iota$ (E12).}
The inventory-expectation sweep is the strongest market-share case in the campaign and
the most statistically expensive. It achieves full pairwise separation for both unemployment
(15 of 15) and market share (15 of 15) at the final step, but at a mean cost of roughly
10\,000 simulations per observable --- four to five times the campaign average. Market share alone
requires around 14\,000 simulations on average, reflecting the fat-tailed distribution of sample
counts needed to resolve the tight precision target ($\delta = 0.0005$) for this observable.

\begin{figure}[h!]
\centering
\includegraphics[width=0.9\textwidth]{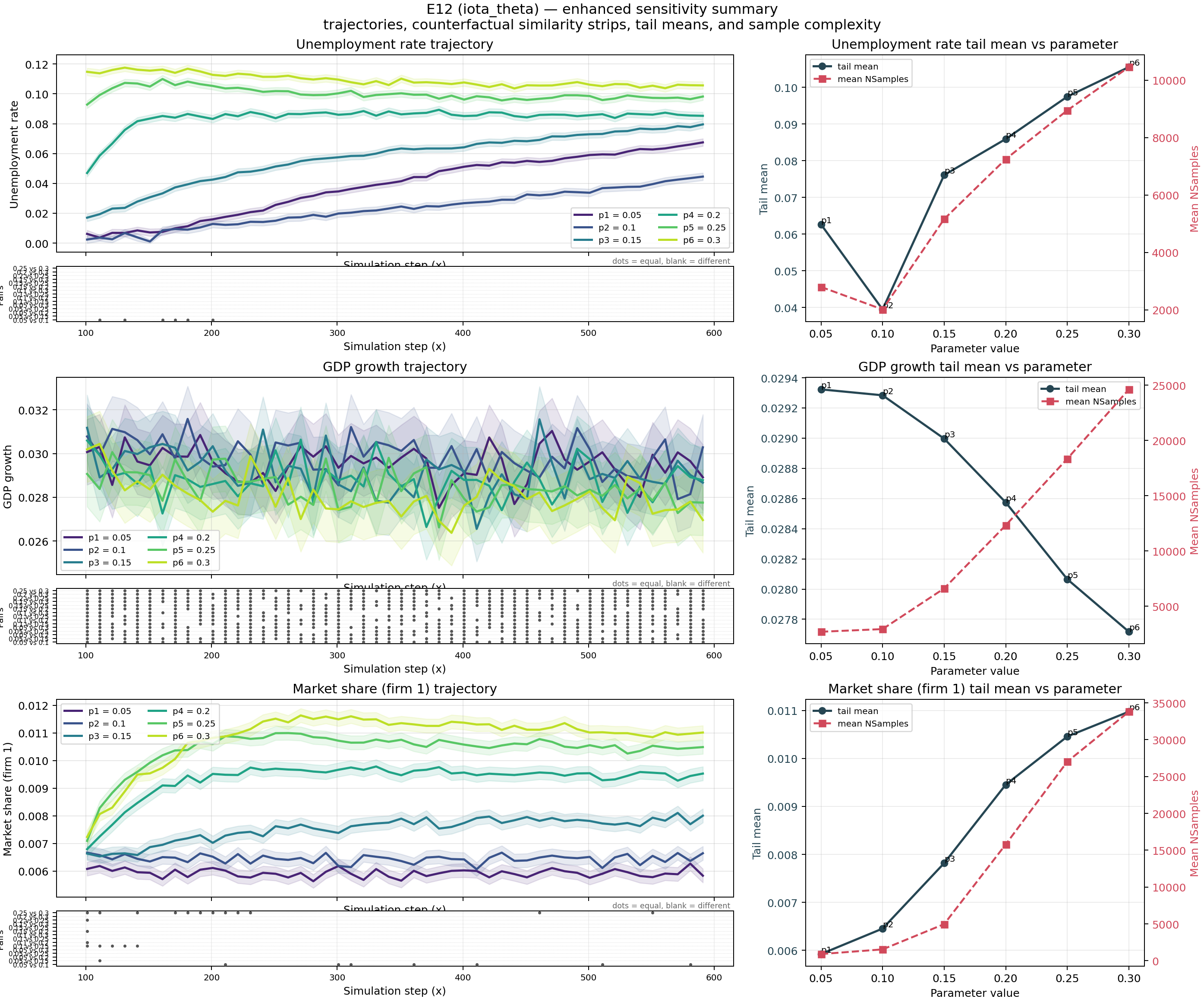}
\caption{Sensitivity summary for `E12' ($\iota$, inventory expectations). This is the
strongest market-share case and the costliest sweep in the campaign.}
\label{fig:e12-iota}
\end{figure}

Figure~\ref{fig:e12-iota} makes the cost-signal trade-off visible: market-share trajectories
separate sharply, whereas GDP growth remains comparatively difficult to resolve despite the much
larger sample budget demanded by the stopping rule. The high statistical cost is itself
informative, because it reveals that the inventory-expectation parameter governs a mechanism
whose micro-level effects require substantially more simulation effort to distinguish
statistically.

\paragraph{Pareto heterogeneity $a$ (E11).}
The Pareto-shape sweep is an outlier in the opposite direction. Despite covering a broad range
of firm-size heterogeneity, it achieves only moderate unemployment separation (6 of 15 final
pairs) and very limited response in the other two observables. Its appendix plot
appears in Appendix~\ref{sec:appendix-figures}.

\section{Discussion}
\label{sec:discussion}

The campaign yields three main types of evidence that are more transparently comparable under common-precision SMC than
under fixed-budget Monte Carlo.

\paragraph{Heterogeneous sensitivity under a common discipline.}
The main methodological payoff is that the sweeps become comparable under a common inferential rule rather than under a
common run budget. Once precision targets are held fixed, some parameter families remain directionally weak whereas
others produce robust and persistent separation. Under fixed-budget Monte Carlo, the same equal-run-budget design would
be less informative because equal run counts need not imply equal evidential strength. SMC turns heterogeneous
simulation costs into part of the empirical record.

\paragraph{Financial parameters dominate; switching parameters are surprisingly weak.}
Among the twelve sweeps, the strongest and most consistent signals come from
macro-financial and structural parameters: credit requirements (`E9'), initial bank markup (`E10'), and
inventory expectations (`E12'). More precisely, `E9' is the statistically strongest macro case,
`E10' the clearest directional win-win case, and `E12' the strongest market-share case. This is
consistent with the ABM literature on financial frictions
and macroeconomic fluctuations~\cite{dosi2013income,dosi2015fiscal,Fagiolo2020InnovationFinanceGrowth}. By contrast, the heuristic switching
parameters `E1'--`E3', which define the meta-level mechanism of the selected K+S
variant~\cite{dosi2020rational}, show weak transient sensitivity under one-parameter sweeps. A plausible
interpretation is that the switching ecology stabilizes quickly enough that isolated changes to $\beta$, $\delta_s$, or
$\eta$ do not substantially alter aggregate trajectories within the explored ranges, or that these parameters matter
jointly rather than marginally. More generally, this contrast is not unexpected in the K+S family: when credit
constraints are binding, firms cut investment and production before differences in expectation rules can fully
propagate to macro outcomes~\cite{dosi2020rational,dosi2013income,dosi2015fiscal}.

\paragraph{Limitations.}
The campaign covers only one-parameter sweeps, so interaction effects are not captured. The baseline parameterization
follows the reference study~\cite{dosi2020rational} rather than an empirical calibration, \texttt{MARKET\_SHARE1}
tracks only one firm and should therefore be read as a micro-level probe rather than as a full concentration measure,
while the chosen $\varepsilon = \delta$ values still encode a judgment about practically meaningful effect sizes; these
choices mainly bound the economic interpretation.

\section{Conclusion}
\label{sec:conclusion}

We have shown that statistical model checking, implemented through MultiVeStA, can serve as a rigorous analysis layer
for a demanding macroeconomic agent-based model. Integration required only a minimal bridge around the existing C++ simulator and produced a reproducible sensitivity campaign with explicit
precision targets, temporal queries, confidence levels, and simulation costs.

Results reveal a clear gradient of sensitivity across parameter families. Financial and structural
parameters produce the strongest transient effects, whereas the switching parameters of the
expectation module show weak marginal sensitivity under one-parameter sweeps. `E9' is the strongest macro-level
statistical case, `E10' the clearest directional win-win case, and `E12' the strongest market-share case.
Unemployment reacts most, with `E4' the clearest policy-side case; GDP growth is harder to move, and the auxiliary market-share probe responds
selectively. In this sense, the campaign refines rather than overturns earlier insights from the K+S family: the prominence of credit-side and
financial mechanisms is recovered here under a transparent common-precision protocol rather than under coarser ad hoc
sensitivity exercises~\cite{dosi2013income,dosi2015fiscal,dosi2020rational}.

More broadly, the paper contributes a reusable workflow for macroeconomic ABMs extendable to richer temporal
properties~\cite{blando2026island}.

\clearpage
\bibliographystyle{splncs04}
\bibliography{references}

\appendix

\section{Summary Tables}
\label{sec:appendix-tables}

This appendix collects the full observable-wise summary tables supporting the compact discussion in
Section~\ref{sec:results}. Tables~\ref{tab:appendix-unempl-summary}--\ref{tab:appendix-market-summary}
report the per-experiment summaries for each observable;
Tables~\ref{tab:appendix-bestpoints}--\ref{tab:appendix-counterfactual} provide the best-tail
configurations, trade-off classifications, and counterfactual separation recaps.

\begin{table}[h!]
\caption{Unemployment summary across the twelve sweeps. Lower tail means are better.}
\label{tab:appendix-unempl-summary}
\centering
\scriptsize
\resizebox{0.9\textwidth}{!}{%
\begin{tabular}{llcccccc}
\toprule
Paper ID & Parameter & Best tail point (min) & Param. value & Tail mean & Final diff. pairs & Tail diff. share & Mean NSamples \\
\midrule
E1 & beta\_switch & $p2$ & 0.2 & 0.039905 & 1/15 & 0.007 & 2267 \\
E2 & delta\_switch & $p2$ & 0.25 & 0.038904 & 0/15 & 0.060 & 2175 \\
E3 & eta\_switch & $p2$ & 0.25 & 0.038045 & 3/15 & 0.160 & 2237 \\
E4 & aliq & $p1$ & 0 & 0.005962 & 15/15 & 0.973 & 3954 \\
E5 & wada\_delta1 & $p1$ & 0.2 & 0.034059 & 12/15 & 0.813 & 2311 \\
E6 & wwtr\_w3 & $p1$ & 0.1 & 0.037651 & 10/15 & 0.620 & 2354 \\
E7 & wstr\_w4 & $p3$ & 1.3 & 0.039482 & 10/15 & 0.620 & 2351 \\
E8 & waa\_laa & $p5$ & 0.8 & 0.039176 & 5/15 & 0.400 & 2262 \\
E9 & tau\_b & $p1$ & 0.04 & 0.011661 & 15/15 & 1.000 & 4348 \\
E10 & mu\_deb\_bankmarkup & $p5$ & 0.5 & 0.005920 & 13/15 & 0.887 & 2403 \\
E11 & paretoa & $p1$ & 0.4 & 0.037643 & 6/15 & 0.267 & 2307 \\
E12 & iota\_theta & $p2$ & 0.1 & 0.039482 & 15/15 & 1.000 & 6105 \\
\bottomrule
\end{tabular}
}
\end{table}

\begin{table}[h!]
\caption{GDP-growth summary across the twelve sweeps. Higher tail means are better.}
\label{tab:appendix-gdp-summary}
\centering
\scriptsize
\resizebox{0.9\textwidth}{!}{%
\begin{tabular}{llcccccc}
\toprule
Paper ID & Parameter & Best tail point (max) & Param. value & Tail mean & Final diff. pairs & Tail diff. share & Mean NSamples \\
\midrule
E1 & beta\_switch & $p5$ & 0.8 & 0.029659 & 1/15 & 0.080 & 2927 \\
E2 & delta\_switch & $p3$ & 0.5 & 0.029476 & 1/15 & 0.047 & 2953 \\
E3 & eta\_switch & $p2$ & 0.25 & 0.029732 & 2/15 & 0.033 & 2940 \\
E4 & aliq & $p1$ & 0 & 0.030278 & 1/15 & 0.267 & 3547 \\
E5 & wada\_delta1 & $p4$ & 0.8 & 0.029971 & 0/15 & 0.060 & 3034 \\
E6 & wwtr\_w3 & $p1$ & 0.1 & 0.029808 & 0/15 & 0.060 & 3084 \\
E7 & wstr\_w4 & $p6$ & 2.5 & 0.029648 & 2/15 & 0.047 & 3078 \\
E8 & waa\_laa & $p3$ & 0.4 & 0.029712 & 1/15 & 0.120 & 2970 \\
E9 & tau\_b & $p1$ & 0.04 & 0.030492 & 7/15 & 0.353 & 3035 \\
E10 & mu\_deb\_bankmarkup & $p4$ & 0.4 & 0.030222 & 0/15 & 0.167 & 3122 \\
E11 & paretoa & $p1$ & 0.4 & 0.029715 & 0/15 & 0.107 & 2944 \\
E12 & iota\_theta & $p1$ & 0.05 & 0.029323 & 2/15 & 0.113 & 11263 \\
\bottomrule
\end{tabular}
}
\end{table}

\begin{table}[h!]
\caption{Market-share summary across the twelve sweeps. Higher tail means are better.}
\label{tab:appendix-market-summary}
\centering
\scriptsize
\resizebox{0.9\textwidth}{!}{%
\begin{tabular}{llcccccc}
\toprule
Paper ID & Parameter & Best tail point (max) & Param. value & Tail mean & Final diff. pairs & Tail diff. share & Mean NSamples \\
\midrule
E1 & beta\_switch & $p1$ & 0 & 0.006463 & 0/15 & 0.000 & 1458 \\
E2 & delta\_switch & $p3$ & 0.5 & 0.006476 & 2/15 & 0.040 & 1494 \\
E3 & eta\_switch & $p1$ & 0 & 0.006464 & 3/15 & 0.040 & 1453 \\
E4 & aliq & $p6$ & 0.25 & 0.006734 & 5/15 & 0.193 & 1852 \\
E5 & wada\_delta1 & $p6$ & 1.2 & 0.006602 & 0/15 & 0.167 & 1536 \\
E6 & wwtr\_w3 & $p6$ & 1 & 0.006643 & 4/15 & 0.140 & 1625 \\
E7 & wstr\_w4 & $p6$ & 2.5 & 0.006663 & 5/15 & 0.267 & 1555 \\
E8 & waa\_laa & $p1$ & 0 & 0.006481 & 2/15 & 0.027 & 1474 \\
E9 & tau\_b & $p3$ & 0.08 & 0.006453 & 0/15 & 0.060 & 1476 \\
E10 & mu\_deb\_bankmarkup & $p2$ & 0.2 & 0.006637 & 7/15 & 0.333 & 1486 \\
E11 & paretoa & $p6$ & 1.5 & 0.006649 & 6/15 & 0.160 & 1514 \\
E12 & iota\_theta & $p6$ & 0.3 & 0.010975 & 15/15 & 0.980 & 14011 \\
\bottomrule
\end{tabular}
}
\end{table}

\begin{table}[h!]
\caption{Best tail configuration for each experiment-observable pair.}
\label{tab:appendix-bestpoints}
\centering
\scriptsize
\resizebox{0.9\textwidth}{!}{%
\begin{tabular}{llccc}
\toprule
Experiment & Observable & Best point & Param. value & Tail mean \\
\midrule
E1 (beta\_switch) & unempl & $p2$ & 0.2 & 0.039905 \\
E1 (beta\_switch) & gdp\_growth & $p5$ & 0.8 & 0.029659 \\
E1 (beta\_switch) & market\_share & $p1$ & 0 & 0.006463 \\
\midrule
E2 (delta\_switch) & unempl & $p2$ & 0.25 & 0.038904 \\
E2 (delta\_switch) & gdp\_growth & $p3$ & 0.5 & 0.029476 \\
E2 (delta\_switch) & market\_share & $p3$ & 0.5 & 0.006476 \\
\midrule
E3 (eta\_switch) & unempl & $p2$ & 0.25 & 0.038045 \\
E3 (eta\_switch) & gdp\_growth & $p2$ & 0.25 & 0.029732 \\
E3 (eta\_switch) & market\_share & $p1$ & 0 & 0.006464 \\
\midrule
E4 (aliq) & unempl & $p1$ & 0 & 0.005962 \\
E4 (aliq) & gdp\_growth & $p1$ & 0 & 0.030278 \\
E4 (aliq) & market\_share & $p6$ & 0.25 & 0.006734 \\
\midrule
E5 (wada\_delta1) & unempl & $p1$ & 0.2 & 0.034059 \\
E5 (wada\_delta1) & gdp\_growth & $p4$ & 0.8 & 0.029971 \\
E5 (wada\_delta1) & market\_share & $p6$ & 1.2 & 0.006602 \\
\midrule
E6 (wwtr\_w3) & unempl & $p1$ & 0.1 & 0.037651 \\
E6 (wwtr\_w3) & gdp\_growth & $p1$ & 0.1 & 0.029808 \\
E6 (wwtr\_w3) & market\_share & $p6$ & 1 & 0.006643 \\
\midrule
E7 (wstr\_w4) & unempl & $p3$ & 1.3 & 0.039482 \\
E7 (wstr\_w4) & gdp\_growth & $p6$ & 2.5 & 0.029648 \\
E7 (wstr\_w4) & market\_share & $p6$ & 2.5 & 0.006663 \\
\midrule
E8 (waa\_laa) & unempl & $p5$ & 0.8 & 0.039176 \\
E8 (waa\_laa) & gdp\_growth & $p3$ & 0.4 & 0.029712 \\
E8 (waa\_laa) & market\_share & $p1$ & 0 & 0.006481 \\
\midrule
E9 (tau\_b) & unempl & $p1$ & 0.04 & 0.011661 \\
E9 (tau\_b) & gdp\_growth & $p1$ & 0.04 & 0.030492 \\
E9 (tau\_b) & market\_share & $p3$ & 0.08 & 0.006453 \\
\midrule
E10 (mu\_deb\_bankmarkup) & unempl & $p5$ & 0.5 & 0.005920 \\
E10 (mu\_deb\_bankmarkup) & gdp\_growth & $p4$ & 0.4 & 0.030222 \\
E10 (mu\_deb\_bankmarkup) & market\_share & $p2$ & 0.2 & 0.006637 \\
\midrule
E11 (paretoa) & unempl & $p1$ & 0.4 & 0.037643 \\
E11 (paretoa) & gdp\_growth & $p1$ & 0.4 & 0.029715 \\
E11 (paretoa) & market\_share & $p6$ & 1.5 & 0.006649 \\
\midrule
E12 (iota\_theta) & unempl & $p2$ & 0.1 & 0.039482 \\
E12 (iota\_theta) & gdp\_growth & $p1$ & 0.05 & 0.029323 \\
E12 (iota\_theta) & market\_share & $p6$ & 0.3 & 0.010975 \\
\bottomrule
\end{tabular}
}
\end{table}

\begin{table}[h!]
\caption{Tail trade-off classification counts against the within-sweep baseline point `p1'.}
\label{tab:appendix-tradeoffs}
\centering
\scriptsize
\resizebox{0.9\textwidth}{!}{%
\begin{tabular}{lccc}
\toprule
Experiment & Win-win & Mixed / trade-off & Lose-lose \\
\midrule
E1 (beta\_switch) & 1 & 4 & 0 \\
E2 (delta\_switch) & 0 & 5 & 0 \\
E3 (eta\_switch) & 1 & 3 & 1 \\
E4 (aliq) & 0 & 0 & 5 \\
E5 (wada\_delta1) & 0 & 3 & 2 \\
E6 (wwtr\_w3) & 0 & 0 & 5 \\
E7 (wstr\_w4) & 0 & 3 & 2 \\
E8 (waa\_laa) & 2 & 3 & 0 \\
E9 (tau\_b) & 0 & 0 & 5 \\
E10 (mu\_deb\_bankmarkup) & 5 & 0 & 0 \\
E11 (paretoa) & 0 & 0 & 5 \\
E12 (iota\_theta) & 0 & 1 & 4 \\
\bottomrule
\end{tabular}
}
\end{table}

\begin{table}[h!]
\caption{Observable-wise counterfactual recap. `Final diff.' counts significant pairwise differences
at the last sampled step; `Tail-majority diff.' counts pairs significant on at least half of the
sampled tail points; `Mean tail diff. share' averages the tail-wise separation over all
sweep-point pairs.}
\label{tab:appendix-counterfactual}
\centering
\scriptsize
\resizebox{0.9\textwidth}{!}{%
\begin{tabular}{llccc}
\toprule
Experiment & Observable & Final diff. & Tail-majority diff. & Mean tail diff. share \\
\midrule
E1 (beta\_switch) & UNEMPL & 1/15 & 0/15 & 0.007 \\
E1 (beta\_switch) & GDP\_GROWTH & 1/15 & 0/15 & 0.080 \\
E1 (beta\_switch) & MARKET\_SHARE1 & 0/15 & 0/15 & 0.000 \\
\midrule
E2 (delta\_switch) & UNEMPL & 0/15 & 0/15 & 0.060 \\
E2 (delta\_switch) & GDP\_GROWTH & 1/15 & 0/15 & 0.047 \\
E2 (delta\_switch) & MARKET\_SHARE1 & 2/15 & 0/15 & 0.040 \\
\midrule
E3 (eta\_switch) & UNEMPL & 3/15 & 2/15 & 0.160 \\
E3 (eta\_switch) & GDP\_GROWTH & 2/15 & 0/15 & 0.033 \\
E3 (eta\_switch) & MARKET\_SHARE1 & 3/15 & 0/15 & 0.040 \\
\midrule
E4 (aliq) & UNEMPL & 15/15 & 15/15 & 0.973 \\
E4 (aliq) & GDP\_GROWTH & 1/15 & 3/15 & 0.267 \\
E4 (aliq) & MARKET\_SHARE1 & 5/15 & 1/15 & 0.193 \\
\midrule
E5 (wada\_delta1) & UNEMPL & 12/15 & 13/15 & 0.813 \\
E5 (wada\_delta1) & GDP\_GROWTH & 0/15 & 0/15 & 0.060 \\
E5 (wada\_delta1) & MARKET\_SHARE1 & 0/15 & 0/15 & 0.167 \\
\midrule
E6 (wwtr\_w3) & UNEMPL & 10/15 & 9/15 & 0.620 \\
E6 (wwtr\_w3) & GDP\_GROWTH & 0/15 & 0/15 & 0.060 \\
E6 (wwtr\_w3) & MARKET\_SHARE1 & 4/15 & 0/15 & 0.140 \\
\midrule
E7 (wstr\_w4) & UNEMPL & 10/15 & 9/15 & 0.620 \\
E7 (wstr\_w4) & GDP\_GROWTH & 2/15 & 0/15 & 0.047 \\
E7 (wstr\_w4) & MARKET\_SHARE1 & 5/15 & 2/15 & 0.267 \\
\midrule
E8 (waa\_laa) & UNEMPL & 5/15 & 6/15 & 0.400 \\
E8 (waa\_laa) & GDP\_GROWTH & 1/15 & 0/15 & 0.120 \\
E8 (waa\_laa) & MARKET\_SHARE1 & 2/15 & 0/15 & 0.027 \\
\midrule
E9 (tau\_b) & UNEMPL & 15/15 & 15/15 & 1.000 \\
E9 (tau\_b) & GDP\_GROWTH & 7/15 & 5/15 & 0.353 \\
E9 (tau\_b) & MARKET\_SHARE1 & 0/15 & 0/15 & 0.060 \\
\midrule
E10 (mu\_deb\_bankmarkup) & UNEMPL & 13/15 & 13/15 & 0.887 \\
E10 (mu\_deb\_bankmarkup) & GDP\_GROWTH & 0/15 & 1/15 & 0.167 \\
E10 (mu\_deb\_bankmarkup) & MARKET\_SHARE1 & 7/15 & 4/15 & 0.333 \\
\midrule
E11 (paretoa) & UNEMPL & 6/15 & 3/15 & 0.267 \\
E11 (paretoa) & GDP\_GROWTH & 0/15 & 0/15 & 0.107 \\
E11 (paretoa) & MARKET\_SHARE1 & 6/15 & 1/15 & 0.160 \\
\midrule
E12 (iota\_theta) & UNEMPL & 15/15 & 15/15 & 1.000 \\
E12 (iota\_theta) & GDP\_GROWTH & 2/15 & 0/15 & 0.113 \\
E12 (iota\_theta) & MARKET\_SHARE1 & 15/15 & 15/15 & 0.980 \\
\bottomrule
\end{tabular}
}
\end{table}

\clearpage
\section{Supplementary Plots}
\label{sec:appendix-figures}

This section collects the supplementary plots for the eight experiments not shown in the
body. Each plot follows the same format as those in Section~\ref{sec:results}: trajectory panels
with confidence bands and counterfactual similarity strips on the left, and tail-mean versus
parameter-value summaries with sample complexity on the right.

\subsection*{Switching Parameters}

The switching sweeps `E1'--`E3' are the weakest block in the campaign. The plots below confirm
visually that the trajectory bands largely overlap for all three observables.

\begin{figure}[h!]
\centering
\includegraphics[width=0.9\textwidth]{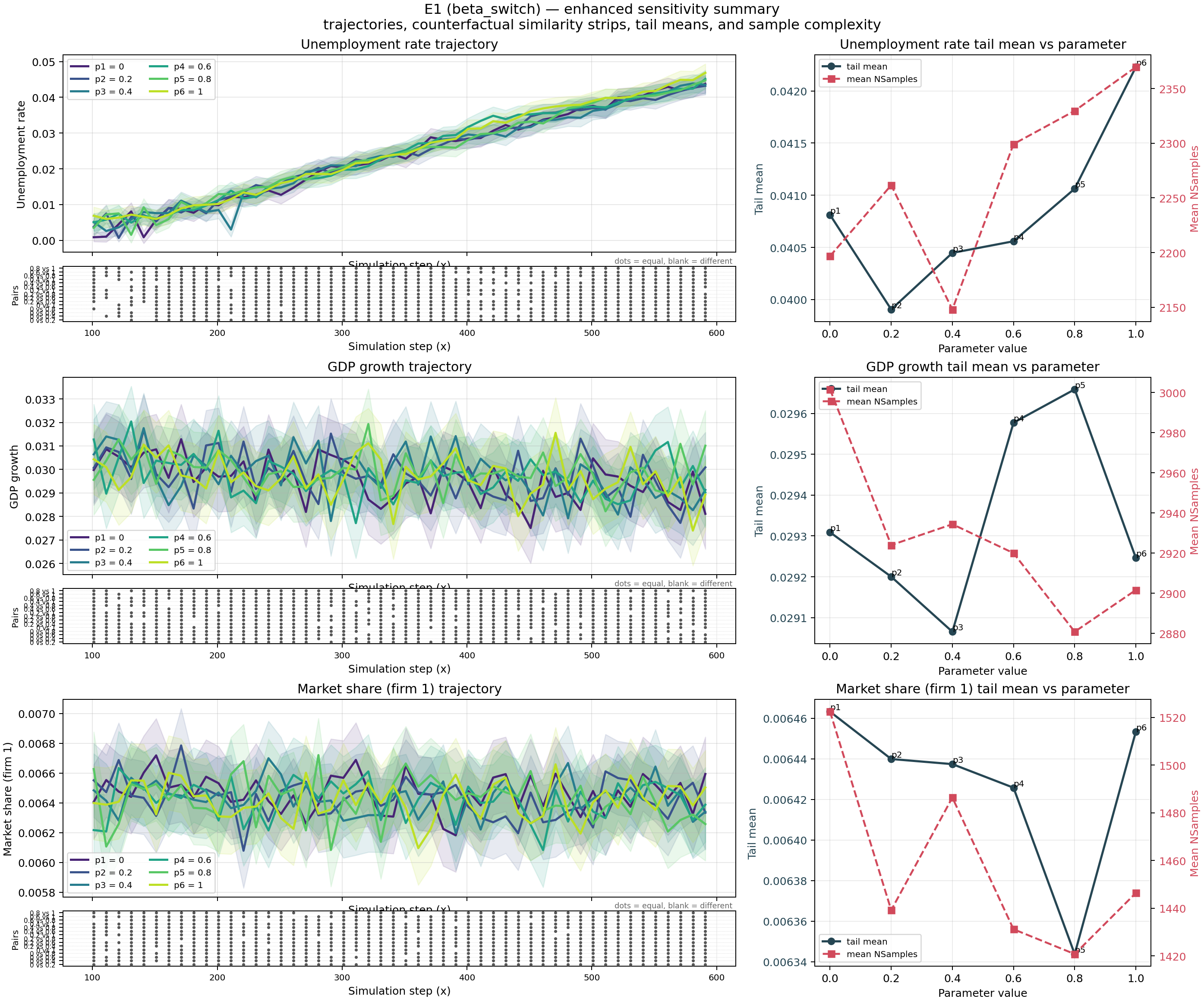}
\caption{Sensitivity summary for `E1' ($\beta$, intensity of choice). Trajectory bands
overlap almost entirely for all three observables, with at most one significant unemployment pair
at the final step.}
\label{fig:e1-beta}
\end{figure}

\begin{figure}[h!]
\centering
\includegraphics[width=0.9\textwidth]{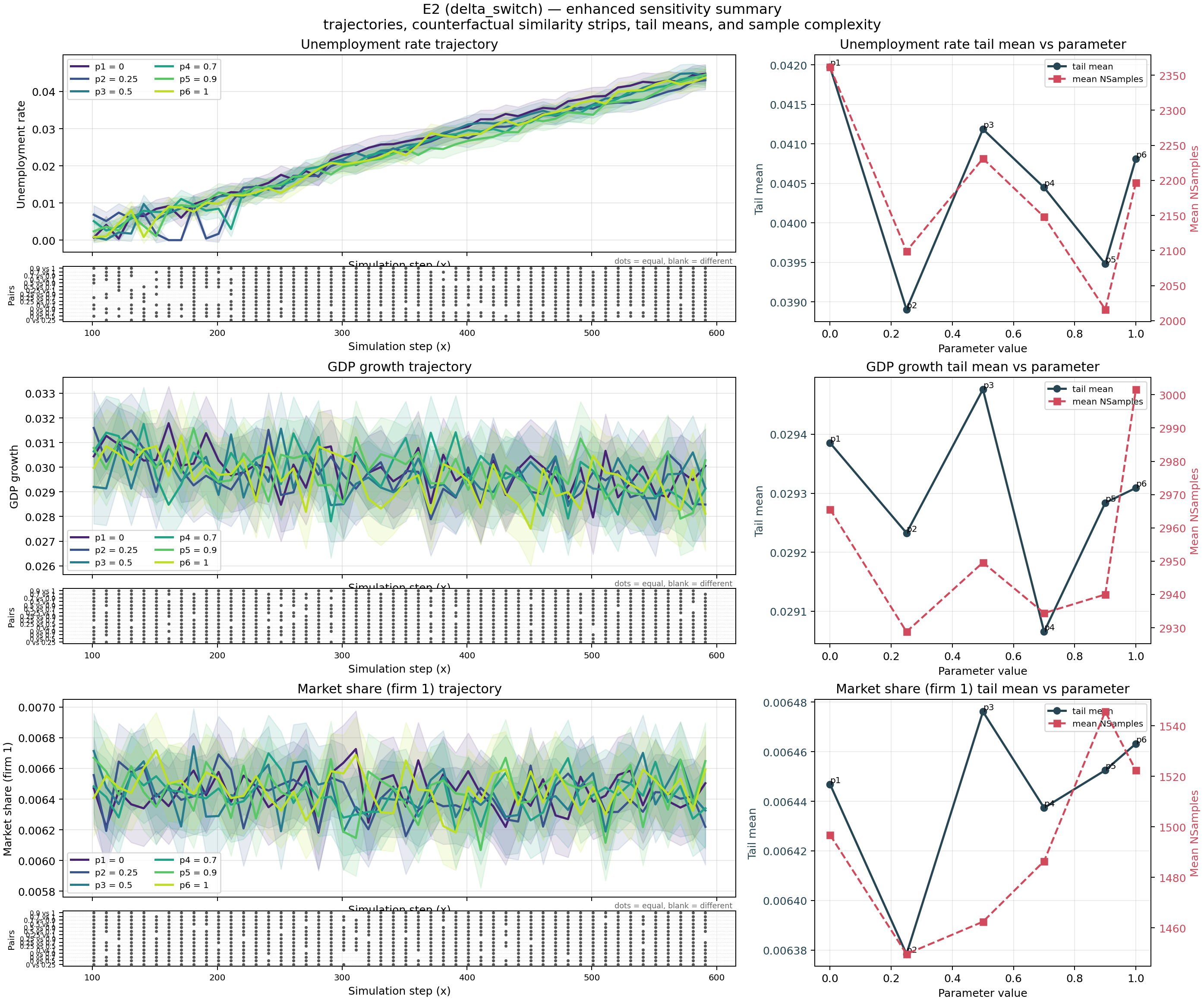}
\caption{Sensitivity summary for `E2' ($\delta_s$, switching inertia). Similar to `E1':
no unemployment pair achieves significance at the final step, and GDP growth and market share
remain flat.}
\label{fig:e2-delta}
\end{figure}

\begin{figure}[h!]
\centering
\includegraphics[width=0.9\textwidth]{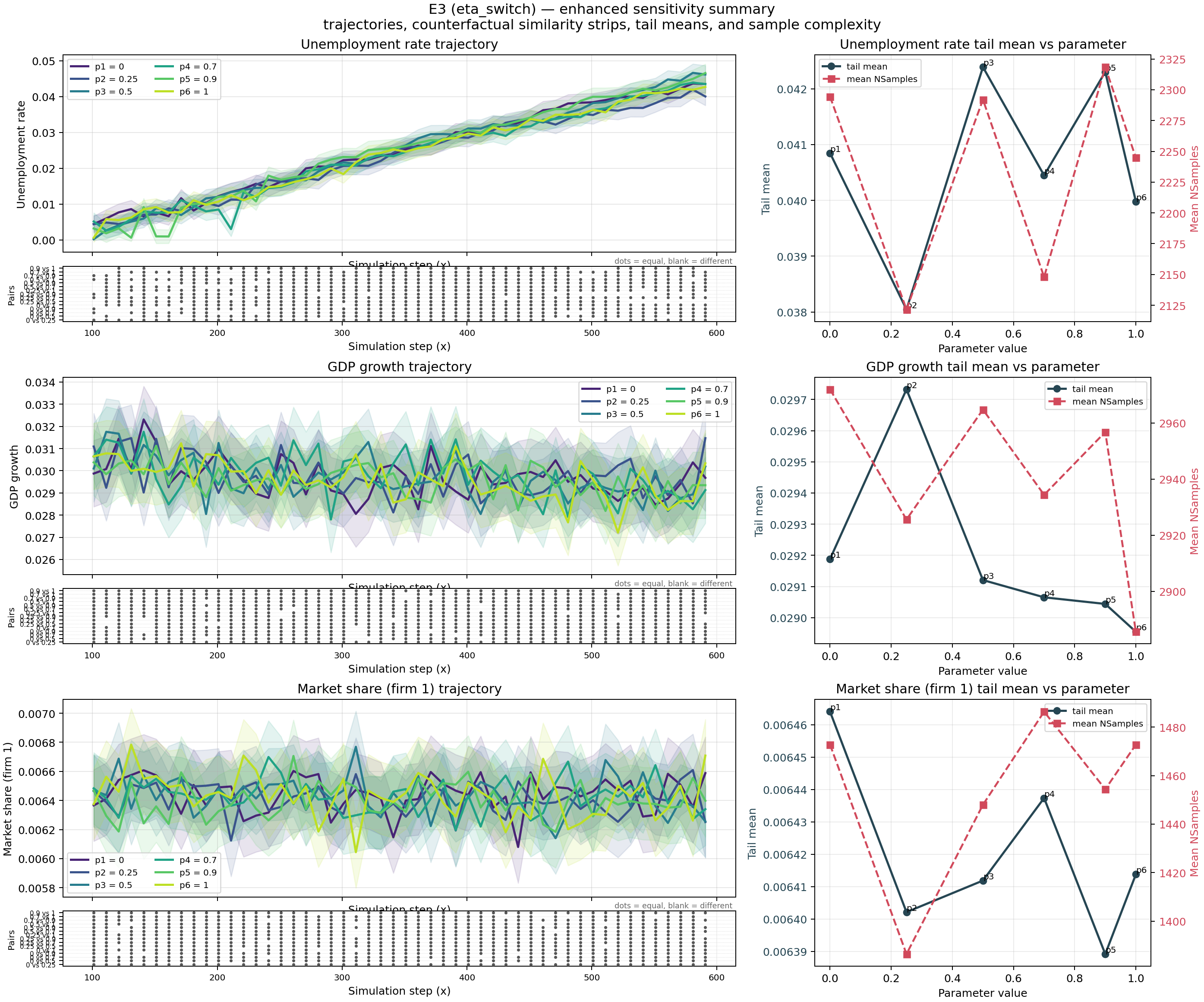}
\caption{Sensitivity summary for `E3' ($\eta$, switching memory). Moderately stronger
than `E1' and `E2' for unemployment (3 of 15 final pairs), but GDP growth and market share remain
largely indistinguishable.}
\label{fig:e3-eta}
\end{figure}

\subsection*{Heuristic Coefficients}

The rule-internal coefficients `E5'--`E8' produce mixed results. Unemployment is often separated,
but GDP growth and market share are rarely moved by these parameters.

\begin{figure}[h!]
\centering
\includegraphics[width=0.9\textwidth]{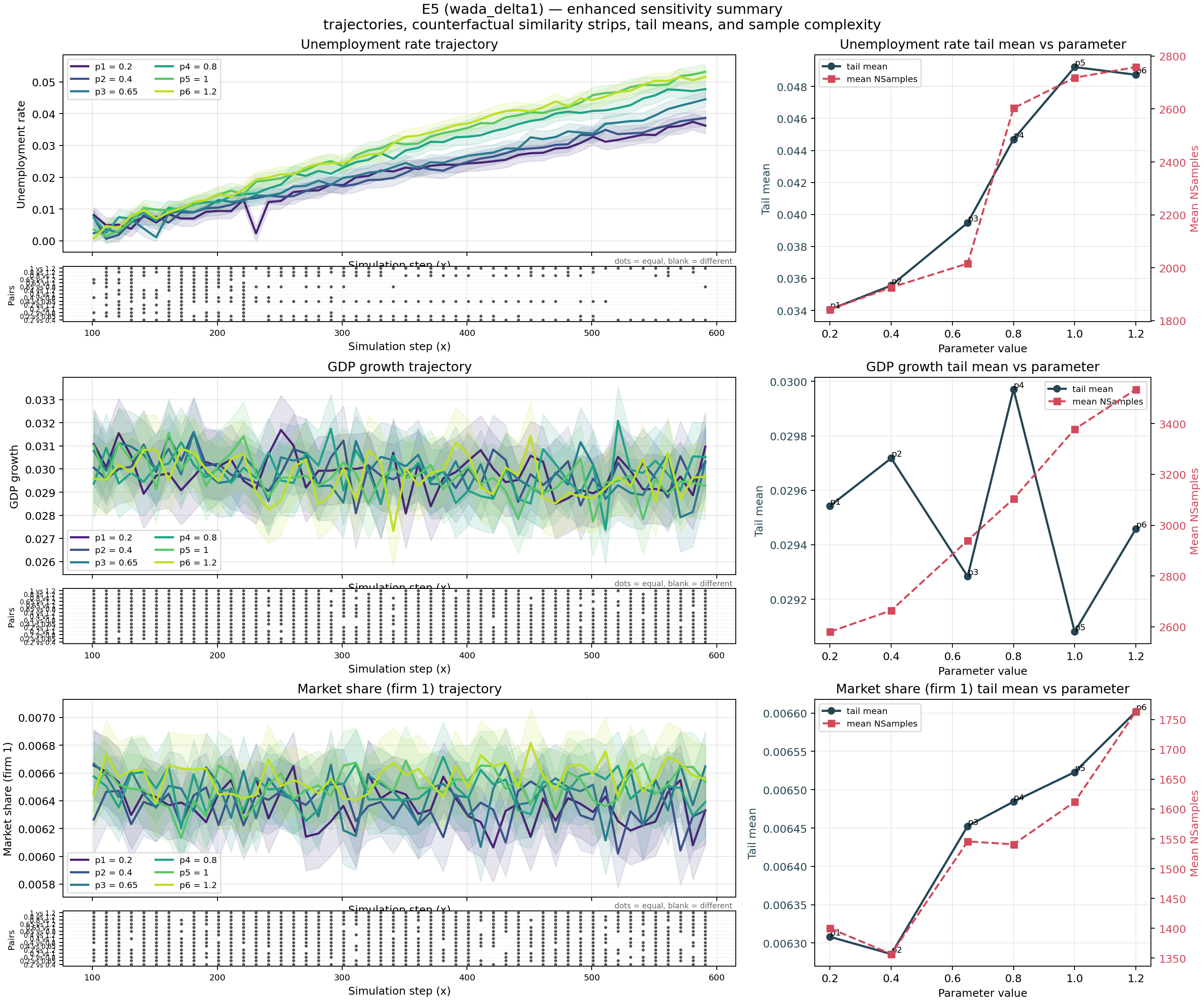}
\caption{Sensitivity summary for `E5' ($\omega_{ada}$, adaptive-expectations coefficient).
Unemployment separation is strong (12 of 15 final pairs), but GDP growth and market share show
limited response.}
\label{fig:e5-wada}
\end{figure}

\begin{figure}[h!]
\centering
\includegraphics[width=0.9\textwidth]{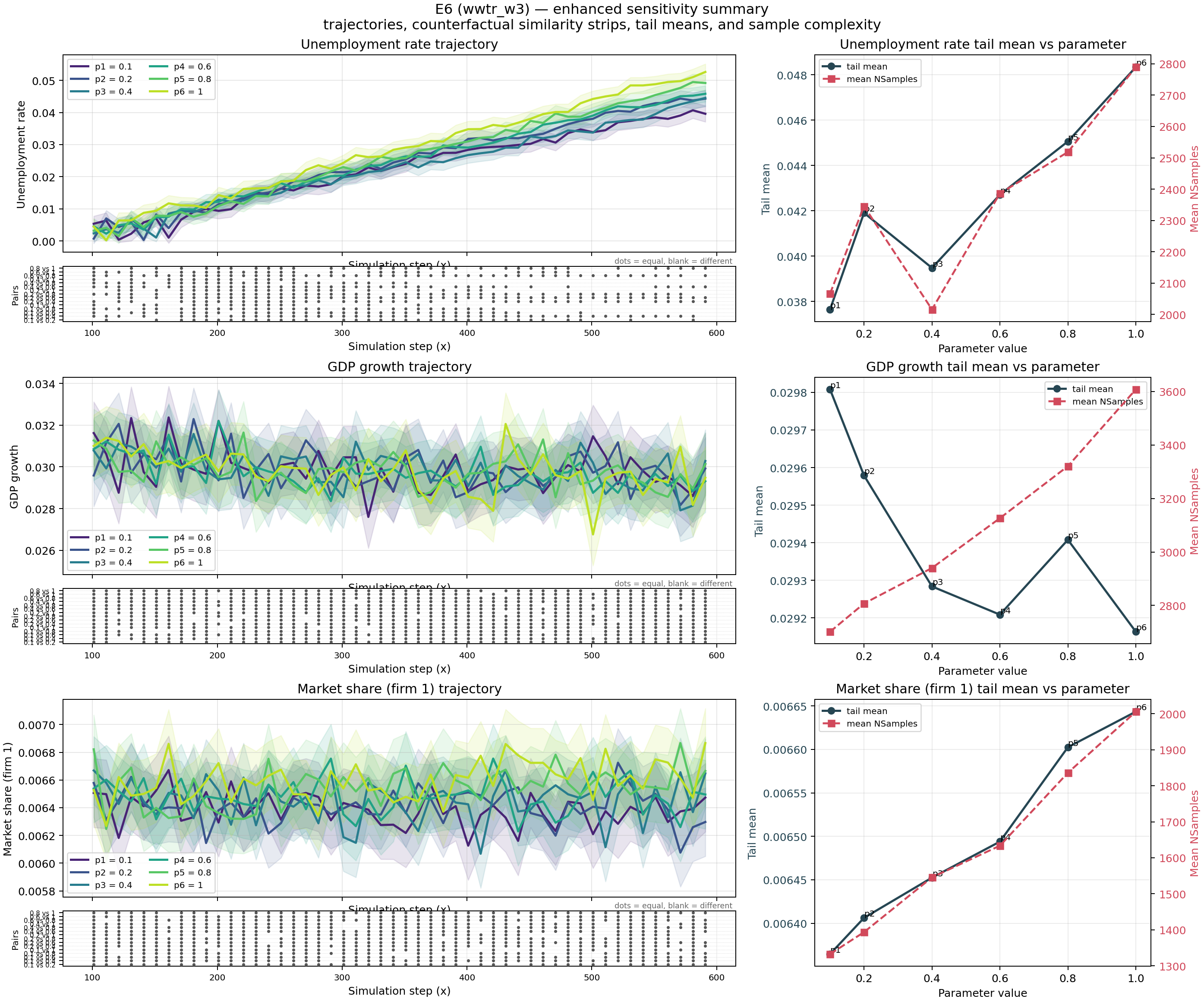}
\caption{Sensitivity summary for `E6' ($\omega_{wtr}$, weak trend-following coefficient).
Unemployment separation is notable (10 of 15 final pairs) but market share responds only weakly
(4 of 15).}
\label{fig:e6-wwtr}
\end{figure}

\begin{figure}[h!]
\centering
\includegraphics[width=0.9\textwidth]{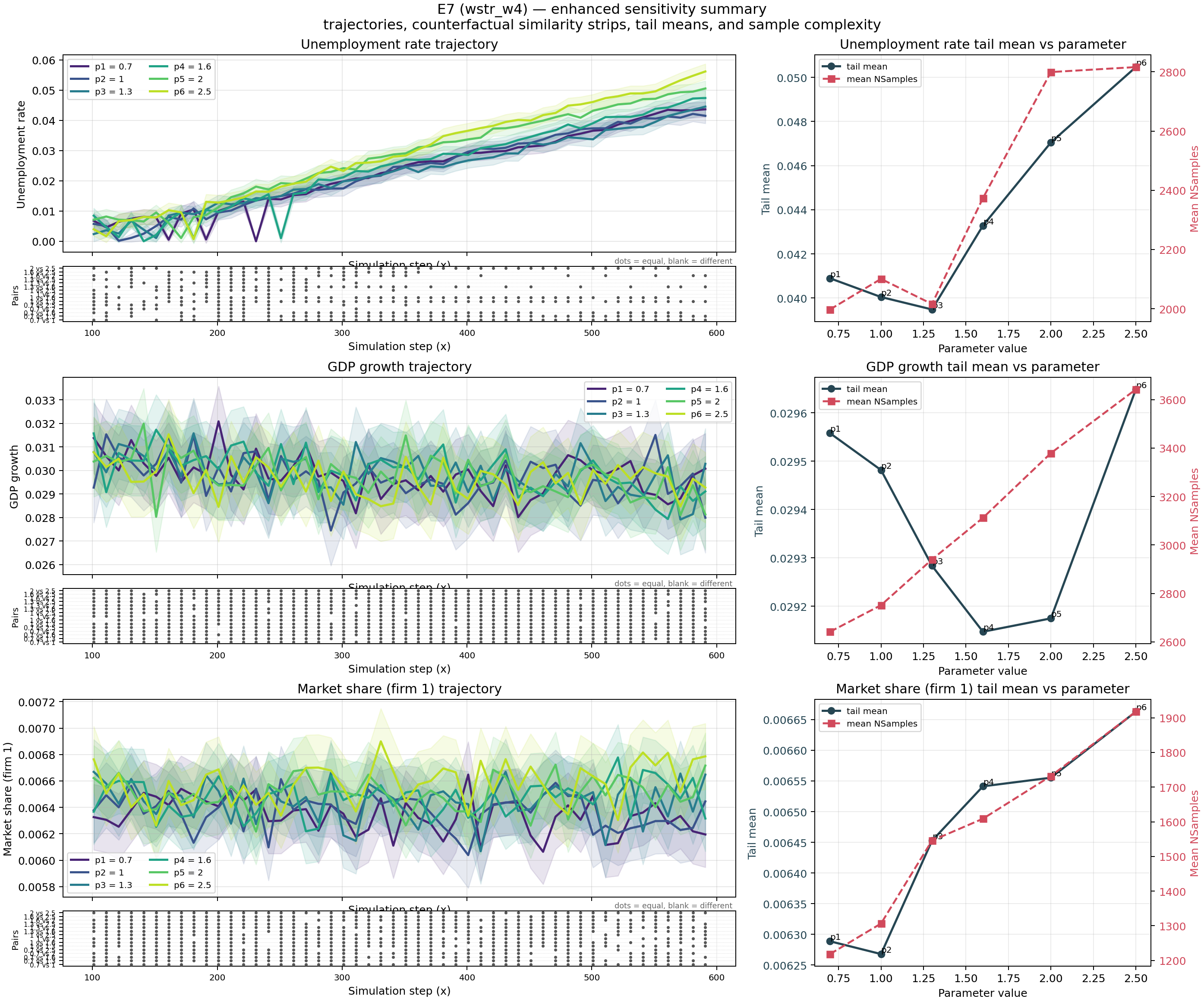}
\caption{Sensitivity summary for `E7' ($\omega_{str}$, strong trend-following coefficient).
Unemployment reaches 10 of 15 final pairs; market share shows moderate response (5 of 15),
making this the third most market-share-reactive sweep after `E12' and `E10'.}
\label{fig:e7-wstr}
\end{figure}

\begin{figure}[h!]
\centering
\includegraphics[width=0.9\textwidth]{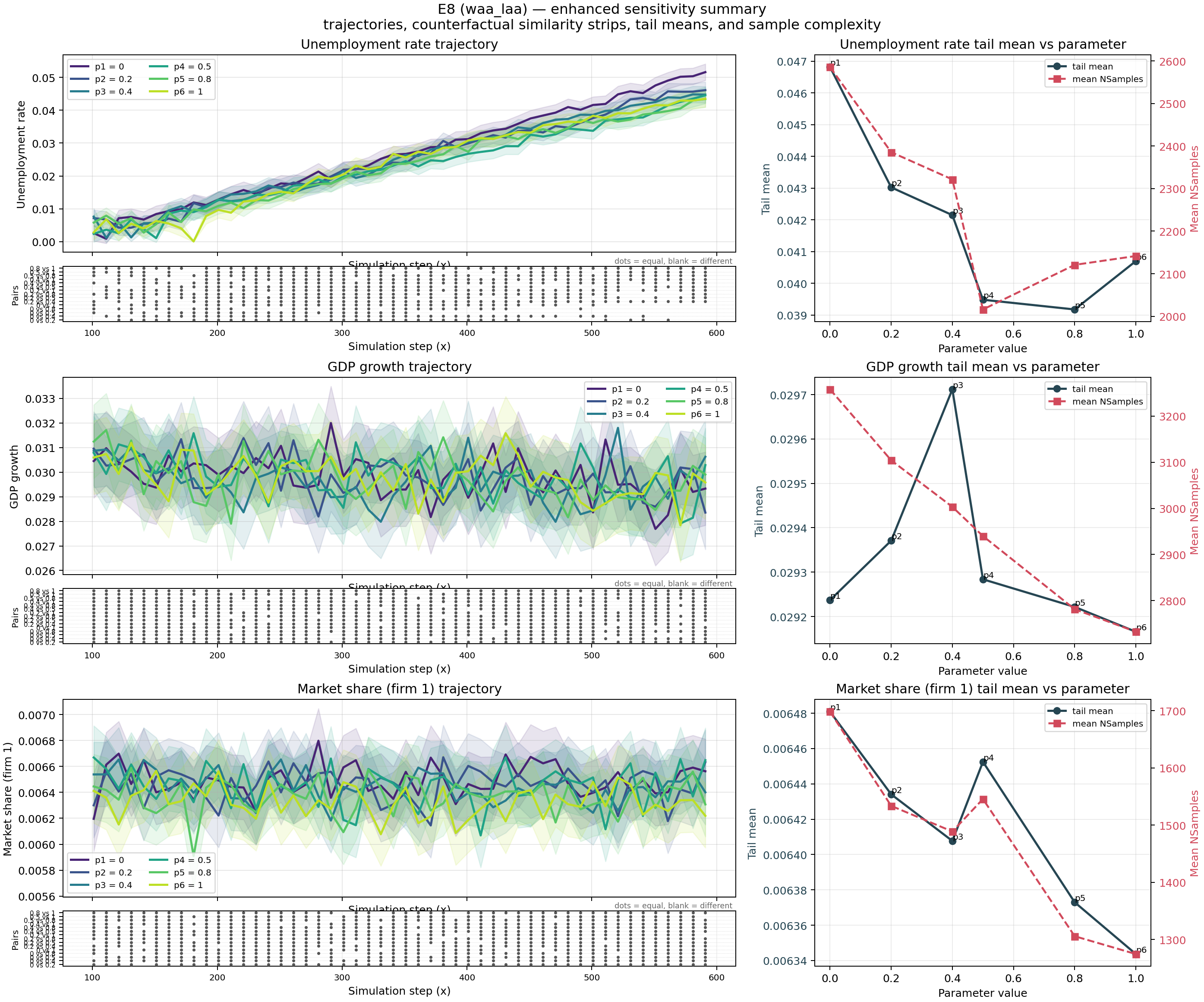}
\caption{Sensitivity summary for `E8' ($\omega_{aa}$, anchor-and-adjustment weight).
Unemployment separation is moderate (5 of 15 final pairs). This is also the only heuristic sweep
with two ``win-win'' points against the baseline.}
\label{fig:e8-waa}
\end{figure}

\subsection*{Financial and Structural Parameters}

The Pareto-heterogeneity sweep `E11' is the only case from this group not shown in the body,
as it produces comparatively weak separation despite covering a broad range of firm-size
heterogeneity.

\begin{figure}[h!]
\centering
\includegraphics[width=0.9\textwidth]{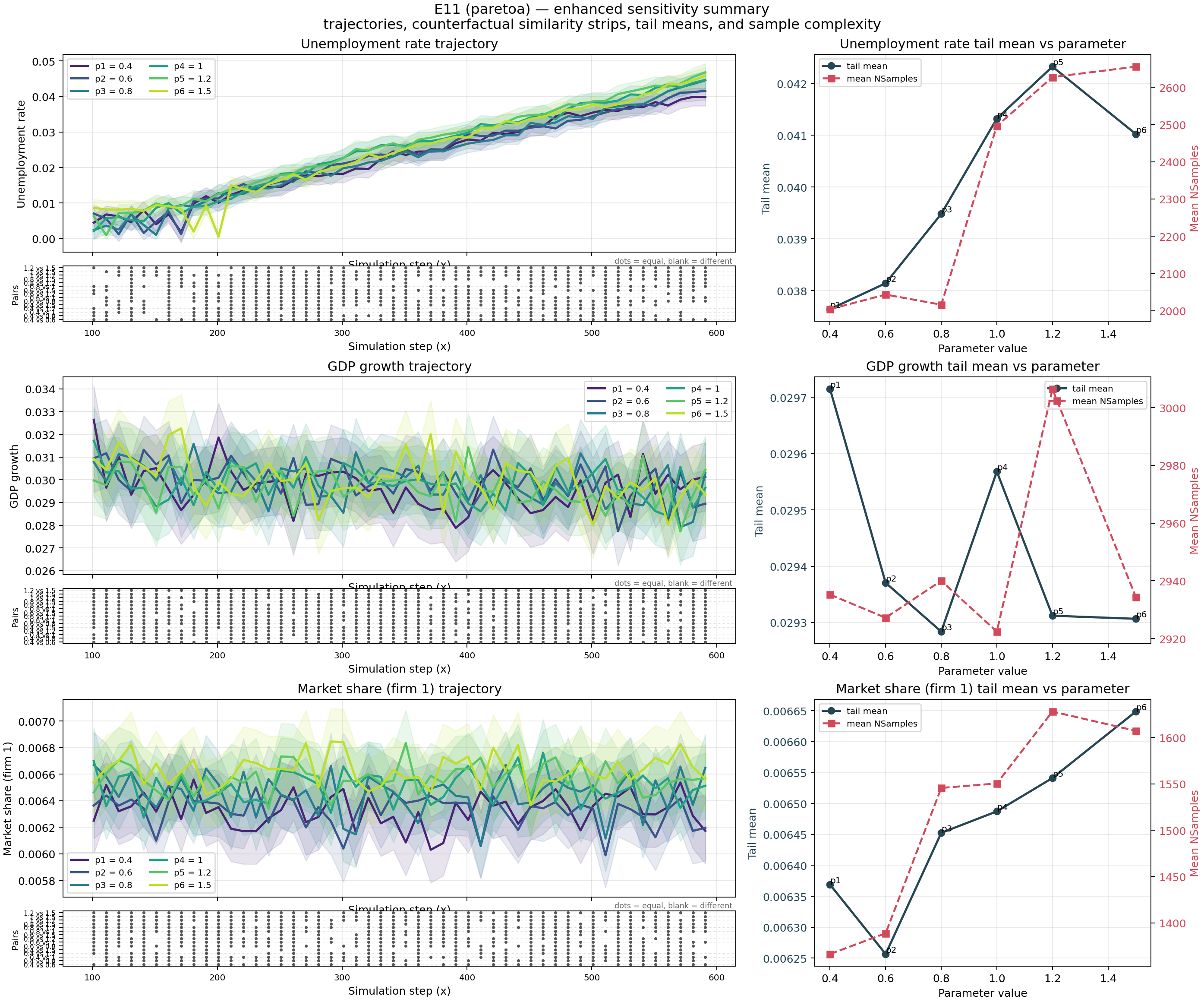}
\caption{Sensitivity summary for `E11' ($a$, Pareto shape). Moderate unemployment
separation (6 of 15 final pairs) but very limited GDP-growth and market-share response, placing
this sweep among the weakest in the financial and structural block.}
\label{fig:e11-paretoa}
\end{figure}

\end{document}